\documentclass[apj]{emulateapj}
\usepackage{natbib}
\usepackage{color}
\usepackage{graphicx}
\bibpunct{(}{)}{;}{a}{}{,}
\begin{document}

   \title{
Localized starbursts in dwarf galaxies produced by 
impact of low metallicity cosmic gas clouds
}
\author{
J. S\'anchez Almeida\altaffilmark{1,2},
B.~G.~Elmegreen\altaffilmark{3}, 
C. Mu\~noz-Tu\~n\'on\altaffilmark{1,2},\\
D.~M.~Elmegreen\altaffilmark{4},
E. P\'erez-Montero\altaffilmark{5}, R.~Amor\'\i n\altaffilmark{6},\\
M.~E.~Filho\altaffilmark{1,2,7,8,9},
Y.~Ascasibar\altaffilmark{10},
P.~Papaderos\altaffilmark{8,9},\\
J.~M.~V\'\i lchez\altaffilmark{5}
}

\altaffiltext{1}{Instituto Astrof\'\i sica de Canarias, 38200 La Laguna, Tenerife, Spain}
\altaffiltext{2}{Departamento de Astrof\'\i sica, Universidad de La Laguna} 
\altaffiltext{3}{IBM Research Division, T.J. Watson Research Center, Yorktown Heights, NY 10598, USA}
\altaffiltext{4}{Department of Physics and Astronomy, Vassar College, Poughkeepsie, NY 12604, USA}
\altaffiltext{5}{Instituto de Astrof\'\i sica de Andaluc\'\i a,  CSIC, Granada, Spain}
\altaffiltext{6}{INAF-Osservatorio Astronomico di Roma, Monte Porzio Catone, Italy}
\altaffiltext{7}{SIM/CENTRA, Lisbon, Portugal}
\altaffiltext{8}{Centro de Astrof\'isica da Universidade do Porto, Porto, Portugal}
\altaffiltext{9}{Instituto de Astrof\'\i sica e Ci\^encias do Espa\c co,  Universidade de Lisboa, Lisboa, Portugal}
\altaffiltext{10}{Universidad Autónoma de Madrid, Madrid, Spain}
\email{jos@iac.es}
\begin{abstract}
Models of galaxy formation predict that gas accretion from the cosmic web is a
primary driver of star formation over cosmic history. Except
in very dense environments where galaxy mergers are also important, 
model galaxies feed from cold streams of gas from the web that 
penetrate their dark matter haloes.
Although these predictions are unambiguous, the observational support has 
been indirect so far. Here we  report spectroscopic evidence for this process in extremely 
metal-poor  galaxies (XMPs) of the local Universe, taking the form of localized 
starbursts associated with gas having low metallicity.
Detailed abundance analyses based on Gran Telescopio Canarias (GTC) optical spectra of 
ten XMPs  show that the galaxy hosts have metallicities around 
60\% solar on average, while the large star-forming regions 
that dominate their integrated light have low metallicities of some 6\% solar. 
Because gas mixes azimuthally in a rotation timescale (a few hundred Myr), 
the observed metallicity 
inhomogeneities are only possible if the metal-poor gas 
fell onto the disk recently. We analyze several possibilities for the origin of the metal-poor 
gas, favoring the metal-poor gas infall predicted by 
numerical models. If this interpretation is correct, XMPs trace the cosmic web gas in their 
surroundings, making them probes to examine its properties.
\end{abstract}



%
%
\section{Motivation}\label{motivation}
Metals are primarily produced by stars. 
Thus, metal-poor objects are chemically unevolved, providing a
gateway to sample physical conditions and processes characteristic of 
early phases in the Universe. XMP galaxies, defined to have less than one tenth
of the metals in the solar composition  \citep{2000A&ARv..10....1K}, 
are rare in the
local Universe \citep[$<0.2\,$\%\ of all galaxies; ][]{2011ApJ...743...77M}. 
They tend to be dwarf galaxies but, unlike other dwarfs \citep{1996ApJ...471..211K},
their chemical composition may not be uniform
\citep[][]{2006A&A...454..119P,2009A&A...503...61I,2011ApJ...739...23L,2013ApJ...767...74S,2014ApJ...783...45S}, 
with the low metallicity only in regions of intense star formation.
Such non-uniformity is particularly revealing because the timescale 
for azimuthal mixing in normal galaxies is short, of the order of a fraction
of the rotational period  \citep[e.g.,][]{2002ApJ...581.1047D,2012ApJ...758...48Y}. 
It would imply that the metal-poor gas in XMPs was recently accreted from 
a nearly pristine cloud, 
providing long-sought evidence for cosmic accretion that has been  
difficult to observe directly \citep[][]{2014A&ARv..22...71S}. 
For example,
gas flows have been observed on the
periphery of distant galaxies using absorption lines in the spectra of
background sources. However, the material is often 
not significantly lower in metals than the galaxy itself, suggesting gas recycled 
from previous outflows, and therefore dynamically detached from the large-scale 
cosmic web \citep{2013ApJ...770..138L,
2013ApJ...779...87C,
2014MNRAS.445.2061L}. 

Here we put on firm observational bases the existence of chemical inhomogeneities 
in most XMPs, thus unequivocally showing their star formation to be feeding
from external metal-poor gas.
Measuring chemical inhomogeneities in XMPs is
technically challenging, requiring spectra of high quality from faint regions. We
used the 10.4\,m GTC\footnote{{\tt http://www.gtc.iac.es/GTChome.php}}  
(Sect.~\ref{observations})
to measure oxygen abundance along the major axes of ten XMP galaxies 
(Table~\ref{table}) with 
a robust variant of the 
direct method \citep[][see Sect.~\ref{metallicity}]{2014MNRAS.441.2663P}. 
In nine out of ten cases, sharp metallicity
drops were found (Sect.~\ref{analysis}).
The origin of the metal-poor gas is analyzed in Sect.~\ref{conclusions},
leaving as the most convincing hypothesis the gas infall from the 
web predicted by numerical models.

\section{Observations}\label{observations}

The ten galaxies in Table~\ref{table} represent  7\,\% of all known 
XMPs when the project began \citep{2011ApJ...743...77M}, 
and they were chosen to be representative of the XMP family. 
They were observed with the long-slit mode of the spectrograph
OSIRIS at GTC. 
We used an intermediate resolution grism which renders
the spectral range needed for abundance analysis in a single exposure 
(from 3700\,\AA\ to 7000\,\AA) with a dispersion of 7.8\,\AA\ per pixel, 
enough to spectrally resolve the required emission lines. The width of the spectrograph 
slit was set to 1\,arcsec, which matches the typical seeing during observation.
Integration times amount to two hours per target. The spectra were reduced 
and calibrated using standard modules of the package IRAF\footnote{{\tt http://iraf.noao.edu/}},
 with the line fluxes computed from the calibrated spectra using custom-made 
IDL routines. 
%
%
%
\begin{deluxetable*}{cccccc}
\tablecaption{Extremely metal poor galaxies selected in this study}
\tablehead{
\colhead{Name\tablenotemark{a}}&
\colhead{$12+\log({\rm O/H})$\tablenotemark{b}}&
\colhead{$\log M_\star$\tablenotemark{c}}&
\colhead{D\tablenotemark{d}}&
\colhead{HII Size\tablenotemark{e}}&
\colhead{$\Delta Z$\tablenotemark{f}}
}
\startdata
J$020549.3-094920.3$&$7.58\pm 0.14$&$8.26\pm 0.48$& $26.4\pm 1.8$&$0.18\pm 0.03$& Y\\
J$030331.3-010947.1$&$7.68\pm 0.13$&$8.33\pm 0.33$& $125\pm 9$&$0.72\pm 0.12$&Y\\
J$031300.0+000612.2$&$7.58\pm 0.17$& $7.67\pm0.31$& $119\pm 8$&$0.35\pm 0.12$&Y\\
J$082555.0+353231.0$&$7.49\pm 0.11$&$6.04\pm 0.03$&$9.6\pm 0.7$&$0.076\pm 0.009$& Y\\
J$094416.6+541134.4$&$7.43\pm 0.15$&$7.05\pm 0.05$ & $23.1\pm 1.6$&$0.16\pm 0.2$&Y\\
J$113202.4+572245.2$&$7.59\pm 0.15$&$7.53\pm 0.48$&$22.5\pm 1.6$&$0.093\pm 0.021$& Y\\
J$114506.3+501802.4$&$7.70\pm 0.10$&$6.71\pm 0.07$& $23.6\pm 1.7$&$0.31\pm 0.02$&N\\
J$115133.3-022221.9$&$7.32\pm 0.15$&$6.7\pm 1.0$&$12.6\pm 0.9$&$0.12\pm 0.01$&Y\\
J$210455.0-003522.0$&$7.33\pm 0.10$&$6.19\pm 0.05$&$21.8\pm 1.6$&$0.094\pm 0.021$&Y\\
J$230210.0+004939.0$&$7.54\pm 0.12$&$8.39\pm 0.33$& $138\pm 10$&$0.47\pm0.13$&Y
\enddata
\tablenotetext{a}{RA and DEC in J2000 coordinates.}
\tablenotetext{b}{Oxygen abundance at the starburst.}
\tablenotetext{c}{Stellar mass of the galaxy from MPA-JHU
given in ${\rm M}_\odot$ units.}
\tablenotetext{d}{Redshift-based galactocentric distance, in Mpc, from NED (NASA/IPAC Extragalactic Database).}
\tablenotetext{e}{Seeing-corrected full-width half-maximum size of the starburst, 
in kpc, as inferred from the H$\alpha$ emission.}
\tablenotetext{f}{Does the galaxy show chemical inhomogeneities? Yes or No. --
Note that the only galaxy without chemical inhomogeneities is also the object
of largest metallicity.\\
\\
\\
}
\label{table}
\end{deluxetable*}

\section{Determination of metallicities, stellar masses, 
and star-formation rates}\label{metallicity}

\begin{figure}
\includegraphics[bb=-180 +10 700 700, width=0.365\textwidth,angle=0]{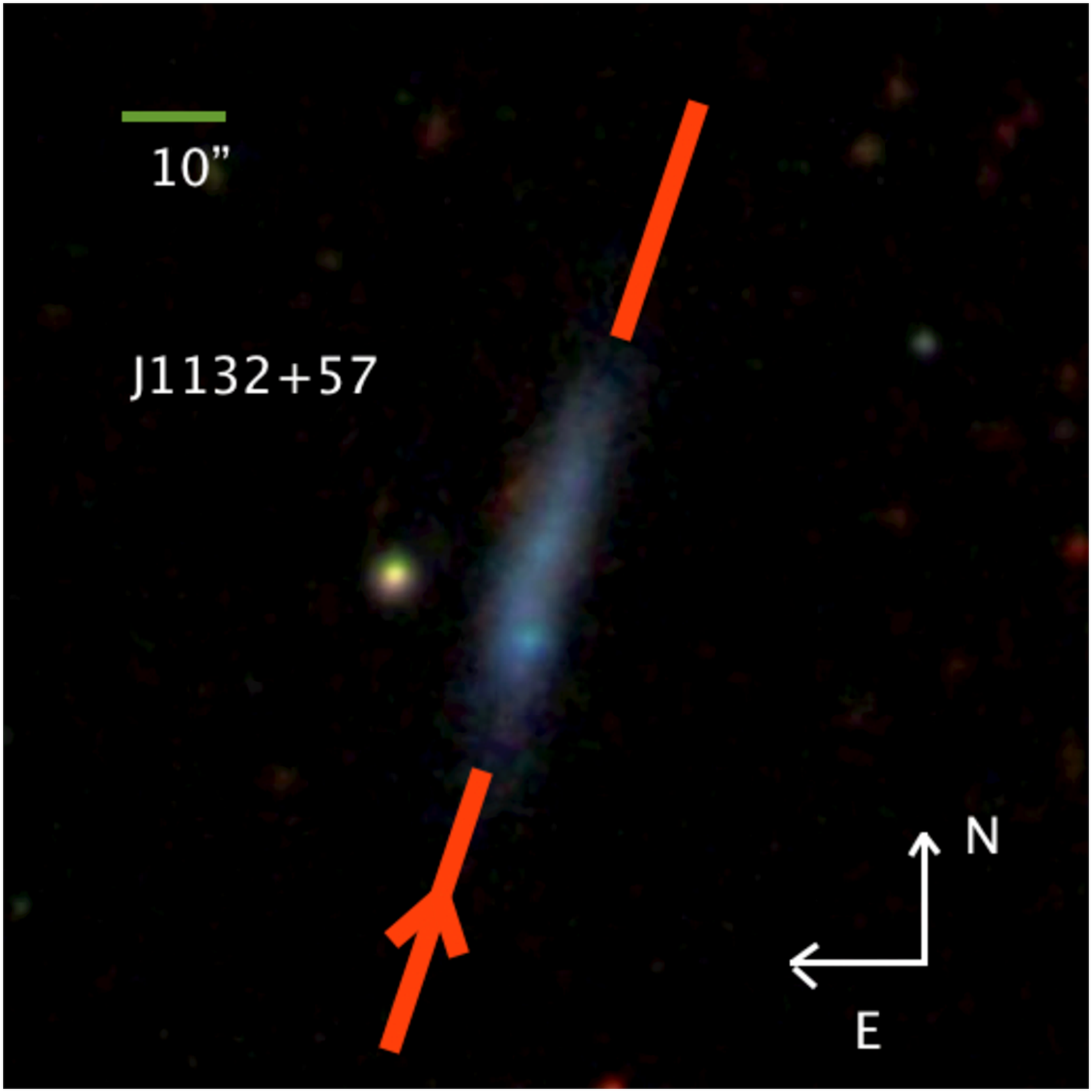}\\
\includegraphics[width=0.5\textwidth,angle=0]{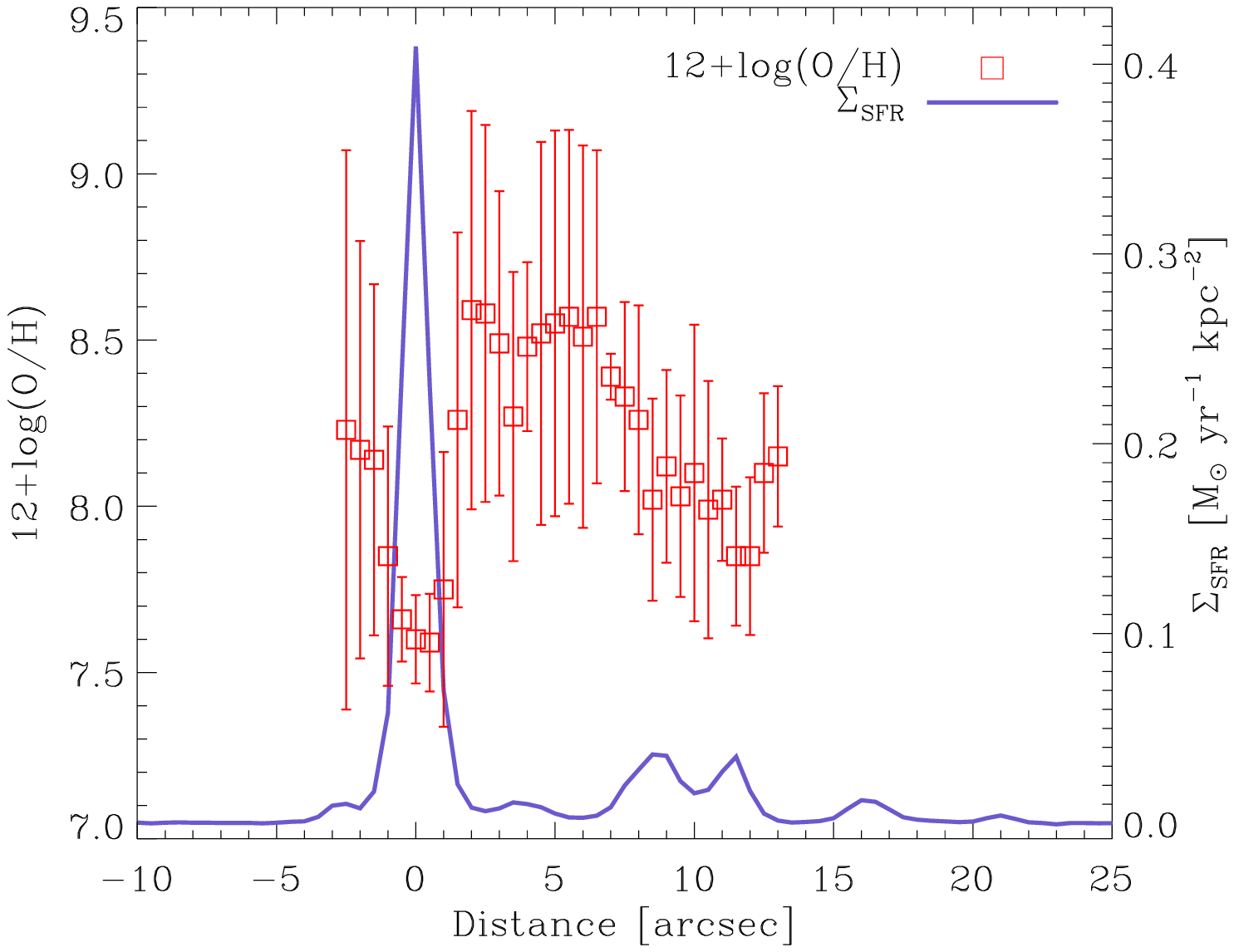}
\caption{
Distribution of oxygen along the major axis of a
typical XMP,  J1132+57.
Top: SDSS image 
with the red bars indicating the position 
of the slit of the spectrograph during observation, 
and the arrow pointing out the sense of growing distances in the bottom panel. 
Bottom: Variation of oxygen abundance 
(red symbols with error bars) and surface SFR (blue solid line)
along the major axis of the galaxy as measured from emission-line 
spectra.
The drop in abundance associated with the peak SFR coincides with the 
brightest knot of the galaxy.
North and east are indicated by white arrows, and the bar on the 
top-left corner corresponds to 10 arcsec on the sky. 
%
}
\label{example}
\end{figure}

\begin{figure}
\includegraphics[bb=-180 +10 700 700, width=0.365\textwidth,angle=0]{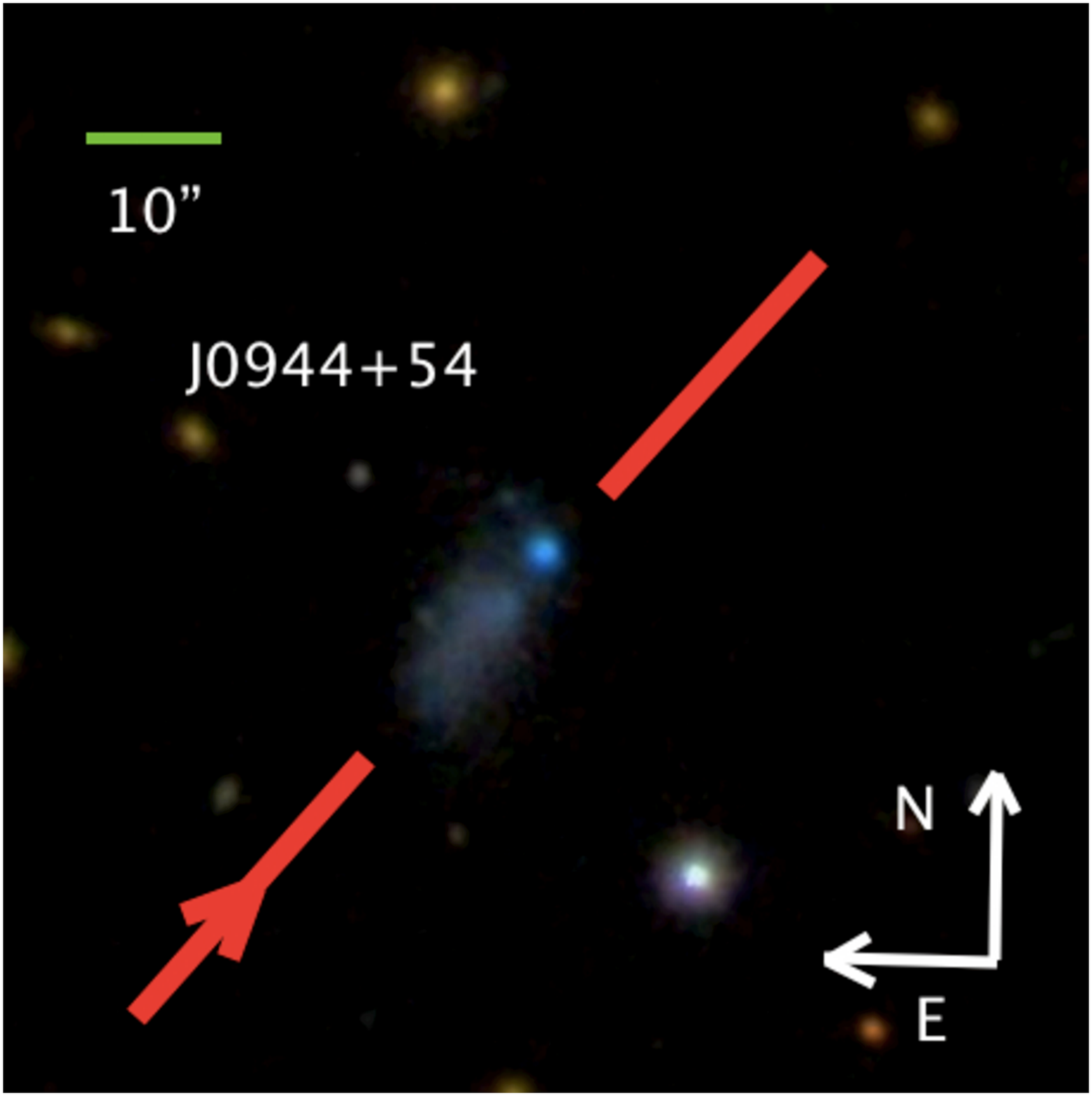}\\
\includegraphics[width=0.5\textwidth,angle=0]{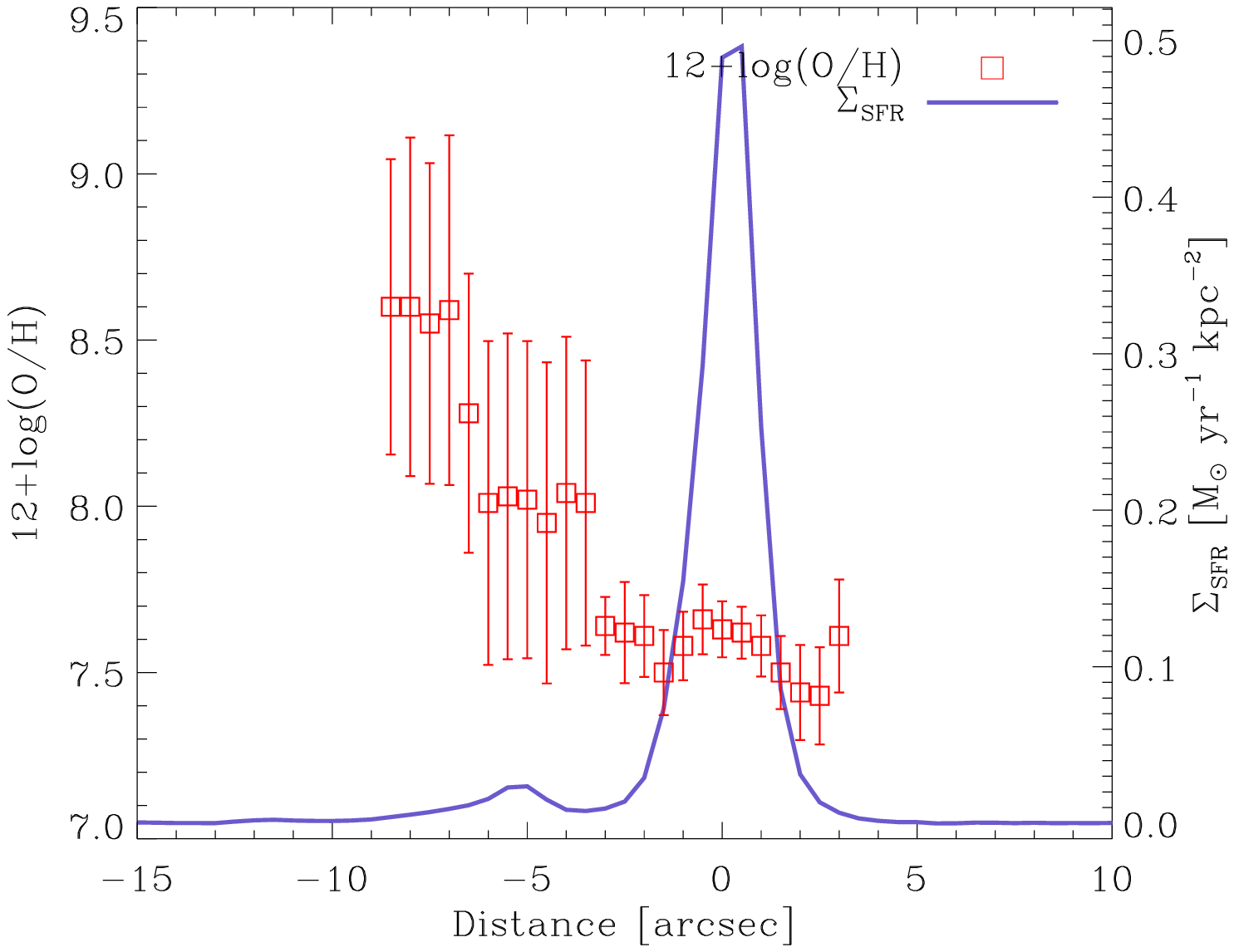}
\caption{
Same as Fig.~\ref{example} for another example, J0944+54.
}
\label{example2}
\end{figure}

HII regions are photoionized by young stars. Their emission-line 
spectrum allows the determination of the oxygen abundance independently of the details of 
the ionization source and the geometry of the nebula
\citep{1974agn..book.....O,2012EAS....54....3S}. 
The spectrum contains a number of collisionally excited
emission lines that empirically provide electron-density and electron-temperature. 
Given temperature and density, atomic properties yield the expected emission 
per ion, therefore, the total flux of a particular emission line directly gives the total
number of ions in the nebula responsible for such emission. 
The case of oxygen is particularly favorable since, in HII regions, oxygen 
is in the form of O$^{+}$ and O$^{++}$, 
and both species produce strong emission lines. The total  oxygen 
abundance results from
adding the contribution of  O$^{+}$ and O$^{++}$. 
A drawback of this technique (the so-called direct method; DM) is the need to 
measure fluxes of weak lines often at the noise level. Unfortunately, this problem 
affects the faint tails of our XMPs (Figs.~\ref{example} and \ref{example2}).  
We circumvent the difficulty using  
HII-CHI-mistry\footnote{{\tt http://www.iaa.es/~epm/HII-CHI-mistry.html}} 
\citep{2014MNRAS.441.2663P},
a procedure that compares the brightest optical  emission lines 
with predictions of a grid of photoionization models covering a wide 
range of possible physical conditions. The resulting metallicity
is calculated as the $\chi^2$-weighted mean of the metallicities in the
models, with the $\chi^2$-weighted standard deviation providing the error.
This method leads to an oxygen abundance which is consistent 
with the DM determination  within 0.07~dex, i.e., within the intrinsic uncertainty of 
the DM \citep{2014MNRAS.441.2663P}. The agreement implies that 
HII-CHI-mistry inherits the good properties of DM, so that 
the geometry and other assumptions needed to build the grid of models
do not influence the inferred oxygen abundance. 
Provided that the emission lines are produced by photoionization,
the agreement between HII-CHI-mistry and DM guarantees that 
irrespectively of the source of ionizing photons,
our abundance determinations along the XMP major axes are reliable
and can be inter-compared.
Photoionization is a good approximation even for the faintest parts of XMPs, 
as attested by the weakness of the line [NII]$\lambda$6583\,\AA . 
In the cases where we have been able to measure it, its flux is less
than a tenth of the H$\alpha$ flux which, according to the  BPT
diagnostic diagram \citep{1981PASP...93....5B}, 
discards ionization by Active Galactic Nuclei, shocks \citep{2013ApJ...774..100K}, 
and evolved stars \citep{2011MNRAS.413.1687C}.
%

The stellar masses of the galaxies given in Table~\ref{table} 
were determined by the MPA-JHU collaboration from galaxy-integrated 
magnitudes \citep{2004MNRAS.351.1151B,2007ApJS..173..267S}. 
Broad-band magnitudes from the SDSS-DR7 database were
compared with a grid of theoretical 
galaxy spectra \citep{2003MNRAS.344.1000B} spanning a large range 
in star-formation histories. 
From the difference between observed and theoretical magnitudes,
a likelihood distribution for the mass of each galaxy is estimated. 
The median and the dispersion of this distribution are used
for the stellar mass and its error.

Star Formation Rates (SFRs) are needed for the discussion of the results
(e.g., Figs.~\ref{example} and \ref{example2}). 
We inferred  them from the observed H$\alpha$ flux as described
by  \citet{1998ARA&A..36..189K}. The underlying hypothesis
is that H$\alpha$ quantifies the number of ionizing photons 
produced by young ($<$~20\,Myr) massive ($>$~10\,M$_\odot$) stars, 
and so, its flux scales with the current SFR.

\section{Results}\label{analysis}

HII-CHI-mistry provides the abundance of oxygen along 
the major axis of the galaxies. In nine out of the ten cases, 
sharp metallicity drops were found (examples are given in 
Figs.~\ref{example} and \ref{example2}, with the only exception
having constant metallicity pointed out in Table~\ref{table}).
Remarkably, the metallicity drops occur coinciding with
starburst regions (the solid lines in Figs.~\ref{example} and \ref{example2}, 
bottom panels).

\begin{figure}
\includegraphics[width=0.5\textwidth,angle=0]{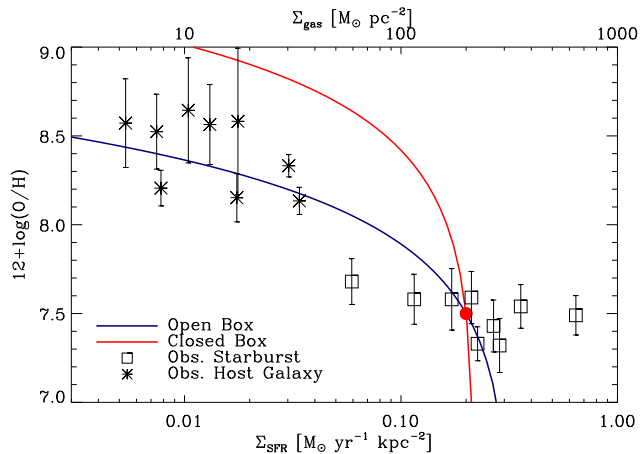}
\caption{
Oxygen abundance of starburst (square symbols) and 
host (asterisks) vs surface SFR for the XMPs with metallicity inhomogeneities.
The axis on top gives the gas surface density for a typical 
gas consumption time scale of one billion years \citep[e.g.,][]{2012ApJ...758...48Y}.
The lines show the chemical evolution
of a starburst at the red bullet as it consumes gas and returns metals;  
the closed-box evolution overproduces oxygen (red line) whereas 
leaking out  80\,\%\ of the gas explains the difference between 
starburst and host (blue line).
}
\label{infoa}
\end{figure}

A summary plot with the metallicity of the starbursts and 
the underlying galaxies is given in Fig.~\ref{infoa}.
The abundance of the starburst was determined as the lowest abundance, whereas
the abundance of the host was computed by averaging all the values in a region of 
2.5\,arcsec around the position with the largest abundance.
 The host galaxies have metallicities higher than the starbursts
by factors of 3 to 10 where their surface SFRs are smaller by factors of 
10 to 100. 
According to the 
empirically established scaling between surface SFR  and gas surface
density
\citep[the so-called Kennicutt-Schmidt law; e.g.,][]{1998ARA&A..36..189K,2012ApJ...758...48Y,2014A&A...564A.121C}, 
the observed starbursts imply large gas reservoirs
with surface densities between $10^2$ and $10^3$~${\rm M_\odot\,pc^{-2}}$ 
(Fig.~\ref{infoa}, upper axis).

According to simple chemical evolution models the starbursts can evolve to 
become the host only if they expel 80\,\%\ of the gas and metals to the 
intergalactic medium (blue solid line in Fig.~\ref{infoa}).
If the evolution proceeds as a closed-box, and it begins from 
pure metal-free gas, the fraction of mass in metals of the 
interstellar medium $Z$ depends only on the gas that 
remains \citep[e.g.,][]{1990MNRAS.246..678E},
so that if a region of metallicity $Z_1$ and gas mass
$\Sigma_{\rm gas1}$ evolves to reach a metallicity
$Z_2$ of gas mass  $\Sigma_{\rm gas2}$, then 
$Z_1-Z_2=Y\,\ln (\Sigma_{\rm gas2}/\Sigma_{\rm gas1})$
with $Y$ the stellar yield, i.e., the mass of new metals eventually 
ejected per unit mass locked into stars. The red solid line in 
Fig.~\ref{infoa} shows the evolution expected for a region at 
the red bullet assuming  a typical Oxygen yield 
\citep[0.004; ][]{2002A&A...390..561M}. 
It renders metallicities too high to explain the 
metallicity of the host galaxy by close-box evolution of the
starbursts. 
If gas is allowed to leak out of the system, the relationship 
between gas mass and metallicity remains formally 
as for the closed-box model, replacing the stellar yield 
with an effective yield $Y_{\rm eff}$ that involves the 
fraction of mass in stars that returns to the interstellar medium $R$,
as well as the so-called mass loading factor $w$, that gives the 
outflow rate in terms of the SFR \citep[e.g.,][]{,2014A&ARv..22...71S}; 
$Y_{\rm eff}=Y\,(1-R)/(1-R+w)$. The blue solid line in Fig.~\ref{infoa}
corresponds to $Y_{\rm eff}=0.2\, Y$, or to $w=3.2$ given $R=0.2$, which 
implies that for each solar mass of gas locked into stars, 
four solar masses of gas flow out of the system.

\begin{figure}
\includegraphics[width=0.5\textwidth,angle=0]{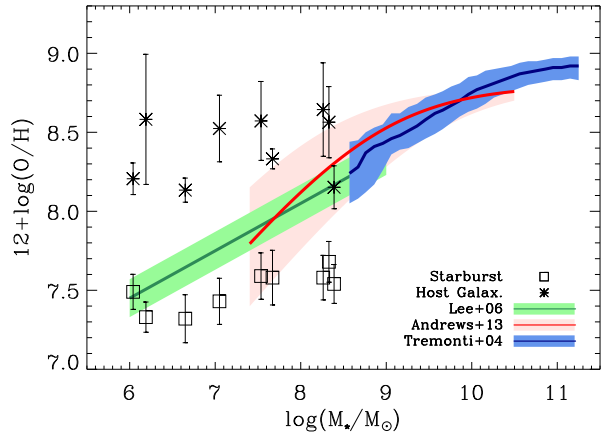}
\caption{
Oxygen abundance of the starburst (square symbols) and 
host (asterisks) vs stellar mass of the galaxy.
We include the relationship observed
in low-mass \citep[green line; ][]{2006ApJ...647..970L}
and high-mass \citep[blue line; ][]{2004ApJ...613..898T} 
local galaxies, the latter re-scaled as described in 
main text.  The colored green and blue regions 
represent the dispersion at a given mass, therefore, XMPs have 
starbursts that are metal-deficient for their masses.  
The dispersion of the metallicity vs mass relation
correlates with SFR \citep{2010MNRAS.408.2115M}
so as to cover the area delimitated by the pink 
region for SFRs going from ten times larger to ten
times smaller than average \citep[][]{2013ApJ...765..140A}.   
Thus, the large difference of SFRs between starburst and
host may naturally account for their different abundances.
}
\label{infob}
\end{figure}

Figure~\ref{infob} shows the scatter plot of metallicity vs galaxy stellar mass
for both the host and the starburst. 
Our objects are outliers of the mass metallicity relationship (MZR)
observed in the local Universe.  The green and the blue lines in Fig.~\ref{infob}  
represent MZRs for the range of low-mass galaxies \citep{2006ApJ...647..970L} 
and high-mass galaxies \citep{2004ApJ...613..898T}, respectively, with the shaded areas 
portraying the observed standard deviation at a given mass. The high-mass metallicities 
have  been decreased by 0.2\,dex to make the strong-line method of the original work 
consistent with the DM scale used here \citep{2008ApJ...681.1183K}. 
The MZR is based on the metallicity of the brightest parts of 
the galaxies therefore, XMPs must be represented by the 
metallicity of their starbursts, which clearly lie below the observed 
MZR (Fig.~\ref{infob}). The scatter at a given mass is known to be related to the 
SFR \citep{2010MNRAS.408.2115M,2010A&A...521L..53L},
so that the higher the SFR the lower the metallicity. The pink area 
in Fig.~\ref{infob} represents the spread
in metallicity to be expected if the SFR changes by  $\pm 10$ times the mean 
SFR given by the red solid line \citep{2013ApJ...765..140A}, and this dependence 
naturally accounts for the large difference in metallicity between starburst and host.
XMPs are going through an active star-formation phase, 
with starbursts of low metallicity. The end of this phase will disclose a galaxy 
having properties close to those of the host, i.e., a stellar mass
similar to the present XMP but with low SFR and high metallicity.

\section{Discussion}\label{conclusions}

The XMPs studied here have an off-center starburst of low 
metallicity but, otherwise, they are rather 
normal disk-like galaxies  with exponential light
profiles, modest rotational velocities, and significant random 
motions \citep{2012ApJ...750...95E,2013ApJ...767...74S}. Under 
these conditions the time-scale for azimuthal gas-mixing is rather 
short (see Sect.~\ref{motivation}), therefore, the metal-poor gas 
associated with the starburst must have arrived to the galaxy recently. 

The mass surface density at the starburst is unusually large for the outer parts 
of dwarf irregular galaxies \citep{2014A&A...564A.121C},
by a factor of $\sim10$. If the excess mass is from a recently 
accreted cosmic cloud, then it mixed with the ambient gas and diluted the ambient 
metallicity by the same factor, 10:1. This dilution accounts for the metallicity drop in the 
starbursts. 
The accretion time is constrained by the localization of the star formation, 
which is not sheared or spiral-like.  The confinement of
the enhanced star formation to local regions several hundred parsecs in diameter
(Table~\ref{table})
implies that the accretion time is shorter than the orbital time. For small galaxies like
these 
the orbital time over one radian is $\sim100$ Myr. Accretion lasting for $\sim100$~Myr 
or less corresponds to an impacting cloud with a vertical extent that is similar to
the radius of the galaxy or several kpc, considering that the infall speed is likely 
comparable to the rotation speed since both are set by the gravitational potential. 
%
%
From these considerations, we conclude the following scenario for XMP galaxies
with locally low metallicities studied here.  Sometime within a period of $\sim100$ Myr, a cosmic
cloud with low metallicity impacted the outer region of a regular dwarf galaxy disk,
compressing it, diluting the original metals, and triggering excess star formation,
all by factors of 10. 
This picture is consistent with the current ideas on how dwarf galaxies grow through
gas accretion 
\citep[Sect.~\ref{motivation} and][]{2008MNRAS.385.2181F,2009Natur.457..451D,2010MNRAS.402.1536S}, 
even though the interaction between gas streams and 
forming disks remains to be properly 
understood \citep{2010MNRAS.404.2151C,2014MNRAS.442.1830V}.

Explanations alternative to the cosmic origin of the gas
cannot be completely ruled out, although they are disfavored
by observations.
Major mergers are not obvious in the optical images, and 
can be further discarded arguing that XMPs reside in low 
density environments, with their large HI envelops 
showing only moderate distortions \citep{2015ApJ...802...82F}.
Minor gas-rich mergers could explain localized 
metallicity drops, provided that the gas in the accreted galaxy
has a metallicity lower than the metallicity at the drop. 
A minor-merger scenario differs from a gas accretion scenario 
only if the merging galaxy has enough stars and dark matter.
The additional mass would generally be much larger than the merging gas mass alone, 
and the distortion on the host would be larger as well. Merging stars should also 
penetrate the host disk, unlike the merging gas, and these stars would orbit 
differently than the host stars, possibly producing star streams in the host halo.
Since neither large HI distortions nor stellar streams are evident in XMPs 
(see, e.g., Figs.~\ref{example} and \ref{example2}), the hypothetical
merging galaxy must have little dark matter and stars and, consequently, it could be 
likened to a gas stream.  On the other hand, model cosmic gas streams are often clumpy, 
with some stars forming in the densest cores \citep[e.g.,][]{2009Natur.457..451D}, 
in which case  the accretion of a clumpy stream is equivalent to
a merger with a small very gas-rich dark-matter-poor satellite.
The difficulty to distinguish a pure gas accretion event from
a gas-rich minor merger can be linked to the problem of 
finding an operational definition of {\em galaxy} at the low-mass 
end of galaxy masses \citep[see][]{2011PASA...28...77F,2012AJ....144...76W}. 
Yet another possibility for feeding XMPs 
could be gas accretion from the galaxy halo 
induced by galactic fountains 
\citep[][]{2008MNRAS.386..935F,2014IAUS..298..228F}. 
The metal-rich gas ejected from the stellar disk acts as a catalyst
when it becomes mixed with the hot metal-poor halo gas.
The mixing catastrophically  decreases 
the halo cooling-time,  and mass from the halo is dragged along 
when the fountain gas falls back onto the disk. 
Several observations disfavor this possibility, though. The 
mass dragged along is only a fraction of the original mass, 
and so, the gas falling in is not very metal-poor.  In addition,
the accreted material is expected to spread out over the disk
rather than being concentrated, and the gaseous haloes in dwarfs are 
not massive, suggesting  the mechanism to be inefficient in our particular case.   

A unidirectional propagation of star-formation may produce a head-tail 
morphology, with decreasing surface brightness and increasing age and 
metallicity from the position of the current starburst. This mechanism was 
proposed by \citet{1998A&A...338...43P} to explain the color gradient in 
the XMP galaxy SBS 0335-052E, and may work for other XMPs as well. The required 
unidirectional propagation would naturally occur if the underlying gaseous structure is
already very prolate. Even if uncommon, cigar-like galaxies are theoretically 
conceivable \citep[e.g.,][]{2003ApJ...584..675B}, and have been observed at high 
redshift \citep[e.g.,][]{2011ApJ...736...92Y}.

The gas-cloud impact scenario explains the lower limit for the metallicities of the XMPs, which is
around 1/50th the solar metallicity \citep[e.g.,][]{2014A&ARv..22...71S}.
The absence of objects with lower metallicity
has challenged interpretation for decades \citep{2011EAS....48...95K}. However, the
threshold is naturally set in this context by the metallicity of the cosmic web gas
at the present time, which is predicted to be around the observed
limit \citep[e.g.,][]{2011MNRAS.418.1796F}. The scenario also explains why XMPs tend to be
cometary \citep{2008A&A...491..113P,2011ApJ...743...77M}. 
The impacting gas clouds
have the largest effect on the outskirts of galaxies where the ambient pressure and
column density are low. For most orientations 
of the galaxy, this gives a cometary shape (like Figs.~\ref{example} and \ref{example2}).

If our scenario is correct, XMPs should not be regarded as pristine galaxies 
but as objects that are still forming in the nearby Universe. 
Their  gas traces the intergalactic medium in their 
immediate surrounding, providing a new method 
to characterize the cosmic web. 

\acknowledgements
%
%
%
%
%
This work has been partly funded by the Spanish Ministry of Economy and 
Competitiveness, projects {\em Estallidos} AYA2013--47742--C04--02--P and
AYA AYA2013--47742--C04--01--P.
RA acknowledges support from European Commission FP7 SPACE project ASTRODEEP (Ref. No: 312725).
PP is supported by FCT through the Investigador FCT Contract 
No. IF/01220/2013 and POPH/FSE (EC) by FEDER funding through the program COMPETE, together
with the project FCOMP-01-0124-FEDER-029170 
(Reference FCT PTDC/FIS-AST/3214/2012), funded by FCT-MEC (PIDDAC) and FEDER (COMPETE).

%
%
\newcommand\pasa{PASA}

\end{document}